\documentclass[10pt,conference]{IEEEtran} 
\usepackage[dvipdfmx]{graphicx}
\usepackage{float}
\usepackage{hyperref}
\usepackage{booktabs}
\usepackage{algorithmic}
\usepackage{textcomp}
\usepackage{xcolor}
\usepackage{soul}
\usepackage{url}
\usepackage{xspace}
\usepackage{multirow}
\usepackage{enumitem}
\usepackage{colortbl}
\usepackage[skins,breakable]{tcolorbox}
\usepackage{framed}
\usepackage{hhline}
\usepackage{threeparttable}
\usepackage{url}
\usepackage{subcaption}
\usepackage{comment}
\usepackage{wrapfig}
\usepackage{subcaption}
\usepackage{caption}
\usepackage{xparse}
\usepackage{listings}
\usepackage{needspace}
\tcbuselibrary{raster,skins}

\def\fig#1{Figure \ref{#1}}
\def\tab#1{Table \ref{#1}}

\def\et{et\ al.\xspace}
\def\eg{e.g.\xspace}
\def\ie{i.e.\xspace}

\definecolor{mygray}{gray}{0.9}
\definecolor{ForestGreen}{HTML}{288c66}
\definecolor{MyBlue}{HTML}{00008b}
\setulcolor{MyBlue}

\newcounter{examplecounter}[subsection]
\renewcommand{\theexamplecounter}{\arabic{section}.\arabic{subsection}.\arabic{examplecounter}}
\newlist{stepitemize}{itemize}{1}
\setlist[stepitemize,1]{leftmargin=1.45cm}

\newlist{phaseitemize}{itemize}{1}
\setlist[phaseitemize,1]{leftmargin=1.70cm}

\NewDocumentCommand{\todo}{o}{
  \fcolorbox{red}{gray!25}{\small\textcolor{red}{\tt{TODO\IfNoValueTF{#1}{}{: #1}}}}
}

\def\summarybox#1#2{
\begin{oframed}
\noindent \textbf{#1:}
#2
\end{oframed}
}

\newcommand{\RQone}{Can mutation analysis results improve the repair success rate of buggy quantum programs?}
\newcommand{\RQtwo}{Can mutation analysis results improve the quality of explanations generated by LLMs?}

\definecolor{keyword}{rgb}{0.67, 0.34, 1.0}  
\definecolor{keyword}{HTML}{F58220}  
\definecolor{comment}{rgb}{0.5, 0.5, 0.5}    
\definecolor{string}{HTML}{EE7800}     
\definecolor{identifier}{HTML}{583F99} 
\definecolor{buggycolor}{HTML}{A00000}
\definecolor{desccolor}{HTML}{0000A0} 
\definecolor{defaultcolor}{HTML}{000000}

\newif\ifhighlight
\highlightfalse 

\lstdefinelanguage{text}{
  basicstyle=\ttfamily\footnotesize,  
  language={},           
  breaklines=true,       
  breakatwhitespace=true, 
  showstringspaces=false 
  breakautoindent=false,  
  breakindent=0pt,         
  escapechar=|,           
}

\newcommand{\lstbg}[3][0pt]{{\fboxsep#1\colorbox{#2}{\strut #3}}}

\lstdefinelanguage{diff}{
  basicstyle=\ttfamily\footnotesize, 
  morecomment=[f][\lstbg{green!20}]+,  
  morecomment=[f][\lstbg{red!20}]-,    
  morecomment=[f][\textit]{@@},     
}

\begin{document}
\title{Leveraging Mutation Analysis for LLM-based Repair of Quantum Programs}
\author{
\IEEEauthorblockN{
   Chihiro Yoshida\IEEEauthorrefmark{1},
   Yuta Ishimoto\IEEEauthorrefmark{2},
   Olivier Nourry\IEEEauthorrefmark{1},\\
   Masanari Kondo\IEEEauthorrefmark{2},
   Makoto Matsushita\IEEEauthorrefmark{1},
   Yasutaka Kamei\IEEEauthorrefmark{2},
   Yoshiki Higo\IEEEauthorrefmark{1}
}
\IEEEauthorblockA{
   \IEEEauthorrefmark{1}The University of Osaka, Osaka, Japan \\
   c-yosida@ist.osaka-u.ac.jp, nourry@ist.osaka-u.ac.jp, matusita@ist.osaka-u.ac.jp, higo@ist.osaka-u.ac.jp
   \\
   \IEEEauthorrefmark{2}Kyushu University, Fukuoka, Japan \\
   ishimoto@posl.ait.kyushu-u.ac.jp, kondo@ait.kyushu-u.ac.jp, kamei@ait.kyushu-u.ac.jp
}
}

\maketitle

\begin{abstract}
In recent years, Automated Program Repair (APR) techniques specifically designed for quantum programs have been proposed.
However, existing approaches often suffer from low repair success rates or poor understandability of the generated patches.
In this study, we construct a framework in which a large language model (LLM) generates code repairs along with a natural language explanation of the applied repairs.
To investigate how the contextual information included in prompts influences APR performance for quantum programs, we design four prompt configurations with different combinations of static information, dynamic information, and mutation analysis results.
Mutation analysis evaluates how small changes to specific parts of a program affect its execution results and provides more detailed dynamic information than simple execution outputs such as stack traces.
Our experimental results show that mutation analysis can provide valuable contextual information for LLM-based APR of quantum programs, improving repair success rates (achieving 94.4\% in our experiment) and in some cases also improving the quality of generated explanations. Our findings point toward new directions for developing APR techniques for quantum programs that enhance both reliability and explainability.
\end{abstract}


\section{Introduction} \label{sec:intro}

Quantum computing has drawn significant attention for its potential to outperform classical machines on specific computational tasks~\cite{nielsen2010quantum}.
To harness the computational power of quantum computers, developers must write \textit{quantum programs} that specify sequences of operations on quantum bits (\textit{qubits}), where each operation is called a \textit{quantum gate}.

Similar to classical (non-quantum) programs, quantum programs also need to be debugged to ensure their correctness~\cite{murillo2025tosem, leite2025tosem}.
However, debugging quantum programs poses distinctive challenges, as it involves quantum-specific concepts such as superposition and entanglement~\cite{shaydulin2020icsew, luo2022saner, zappin2025icse}.
Indeed, these quantum-specific concepts lead to unique software engineering challenges, such as quantum-specific technical debt~\cite{openja2022jss, ishimoto2024apsec} and code smells~\cite{chen2023icse}.

To assist developers in debugging quantum programs, researchers have proposed \textit{automated program repair (APR)} techniques specifically designed for quantum software~\cite{li2024tosem, guo2024qse, tan2025fse}.
The goal of APR is to generate patches that modify buggy programs to pass all test cases in a given test suite~\cite{le2019commacm}.
Guo~\et~\cite{guo2024qse} investigated the capability of ChatGPT~\cite{chatgpt} to repair quantum programs.
Their results showed that ChatGPT could fix only 17\% of complex bugs that involve implementing quantum algorithms.
Tan~\et~\cite{tan2025fse} and Li~\et~\cite{li2024tosem} proposed APR techniques that generate quantum gates as repair patches and insert them into quantum programs.
The current state-of-the-art approach, HornBro~\cite{tan2025fse}, has been shown to fix more bugs than ChatGPT-based APR methods. 
However, this improvement comes at the cost of increased program complexity due to the insertion of additional gates. In one instance, HornBro added up to 249 additional gates while fixing a buggy program, negatively impacting the understandability and maintainability of the repaired code~\cite{chen2023icse}.
In summary, existing APR techniques for quantum programs suffer from either (1) low repair success rates or (2) poor understandability and maintainability of the repaired programs.

In this study, we aim to improve APR for quantum programs by incorporating insights from \textit{mutation analysis} into a large language model (LLM)-assisted repair process.
Mutation analysis evaluates how small changes to specific parts of a program affect its execution results.
For quantum programs, mutation analysis combines classical mutation operators applicable to classical programs (\eg, replacing arithmetic operators) with quantum mutation operations (\eg, adding and deleting quantum gates)~\cite{fortunato2022tqe}.
This approach has demonstrated its effectiveness for testing~\cite{fortunato2022tqe,ali2021icst,mendiluze2021ase,mendiluze2025ese} and fault localization~\cite{ishimoto2025arxiv}.
In our experiments, each program is first analyzed using mutation analysis. 
The LLM is then given the buggy quantum program, its stack trace, and the mutation analysis results as input to generate \textit{fixed code} along with a \textit{natural language explanation} of the applied repair.
Generating a natural language explanation alongside the fixed code helps developers understand the intent and rationale behind the repair, which cannot be easily inferred from the fixed code alone.
We hypothesize that mutation analysis results can improve the repair success rate and help the model produce richer explanations, as they provide more detailed dynamic information than simple execution outputs, such as stack traces.
The research questions (RQs) are as follows:
\begin{enumerate}[label=\textit{RQ\arabic*}, leftmargin=*, align=left]
    \item \textit{\textbf{(Repair success rate)} \RQone}
    \item \textit{\textbf{(Quality of explanation)} \RQtwo}
\end{enumerate}

Our experimental setup includes 18 real-world buggy quantum programs from the Bugs4Q dataset~\cite{zhao2023jss}.
All of these programs are written in Python using Qiskit~\cite{qiskit}, a widely used library for quantum programming.
Throughout the experiment, we use GPT-5~\cite{gpt5}, a state-of-the-art LLM, as the base model for generating the program repairs.
For mutation analysis, we employ QMutPy~\cite{fortunato2022tqe}, a mutation testing framework designed for quantum programs.
We compared four prompt configurations to analyze how different contextual information affects repair performance.
Our main contributions are as follows:
\begin{itemize}[topsep=4pt, partopsep=0pt, parsep=1pt, itemsep=0pt]
    \item We present the first empirical evidence that mutation analysis can serve as a valuable signal for improving LLM-based APR for quantum programs.
    \item We demonstrate that incorporating both dynamic runtime information and mutation analysis results into the prompt can yield a higher repair success rate.
    \item We show that incorporating mutation analysis results can improve the quality of LLM-generated explanations by making positional descriptions more accurate, complete, and concise.
\end{itemize}

\section{Related Work} \label{sec:related}

\subsection{Automated Program Repair for Classical Programs}
Among the APR techniques, learning-based repairs~\cite{zhang2023tosem} have recently gained attention, driven by rapid advances in machine learning, particularly LLMs.
Numerous studies for LLM-based repairs highlight the importance of the contextual information included in the prompts~\cite{jin2023fse, xia2024issta, bouzenia2025icse, ehsani2025arxiv}.
For example, InferFix~\cite{jin2023fse} enhances prompts with static analysis results and bug location, enabling the LLM to generate repairs with explicit awareness of the bug type and its position in the code.
Ehsani~\et~\cite{ehsani2025arxiv} enhance prompts by hierarchically injecting bug-, repository-, and project-level knowledge to provide broader and more relevant repair context.

Our work follows a similar line of research in investigating the impact of prompts on the performance of LLM-based APR.
The novelty of our approach lies in augmenting prompts with mutation analysis results that are effective for testing~\cite{fortunato2022tqe,ali2021icst,mendiluze2021ase,mendiluze2025ese} and fault localization~\cite{ishimoto2025arxiv} for quantum programs.
To the best of our knowledge, no prior work has leveraged mutation analysis to enhance LLM-based APR of either quantum or classical programs.

\subsection{Automated Program Repair for Quantum Programs}
Fixing bugs in quantum programs is challenging because it requires domain-specific knowledge of quantum computing~\cite{shaydulin2020icsew, luo2022saner, zappin2025icse}.  
In a previous study, Guo~\et~\cite{guo2024qse} used ChatGPT~\cite{chatgpt} to explore APR for quantum programs. Their results showed that while ChatGPT successfully repaired 97\% of classical (non-quantum) bugs in their dataset, it only managed to repair 17\% of quantum bugs.
Li~\et~\cite{li2024tosem} proposed UnitAR, which repairs quantum programs by automatically generating unitary operations. 
Tan~\et~\cite{tan2025fse} proposed HornBro, which removes faulty gates responsible for the bug and synthesizes new ones to produce the correct behavior.
Both UnitAR and HornBro can be regarded as \textit{synthesis-based repair} techniques because they generate and insert new gates to fix buggy programs.  
We chose an LLM-based repair approach for the following reasons:
\begin{enumerate}
    \item Unlike synthesis-based approaches, which are limited to gate-related bugs, LLM-based approaches are more flexible and can handle a wider range of bugs such as API-related bugs, which are also commonly found in real-world quantum programs~\cite{luo2022saner}.
    \item Synthesis-based approaches increase the number of gates in a quantum program, thereby increasing its complexity.
    Avoiding excessive gate counts is recommended as a best practice by Google Quantum AI~\cite{google_best_practice} and has also been identified as a type of quantum code smell~\cite{chen2023icse}.
\end{enumerate}
Our approach not only instructs the LLM to generate minimal repairs but also to provide natural language explanations for them.
Unlike other APR techniques for quantum programs, this allows developers to examine and validate the rationale and intent behind each change.
\section{Study Design} \label{sec:study-design}
\begin{figure}[t]
    \centering
    \includegraphics[width=0.90\linewidth]{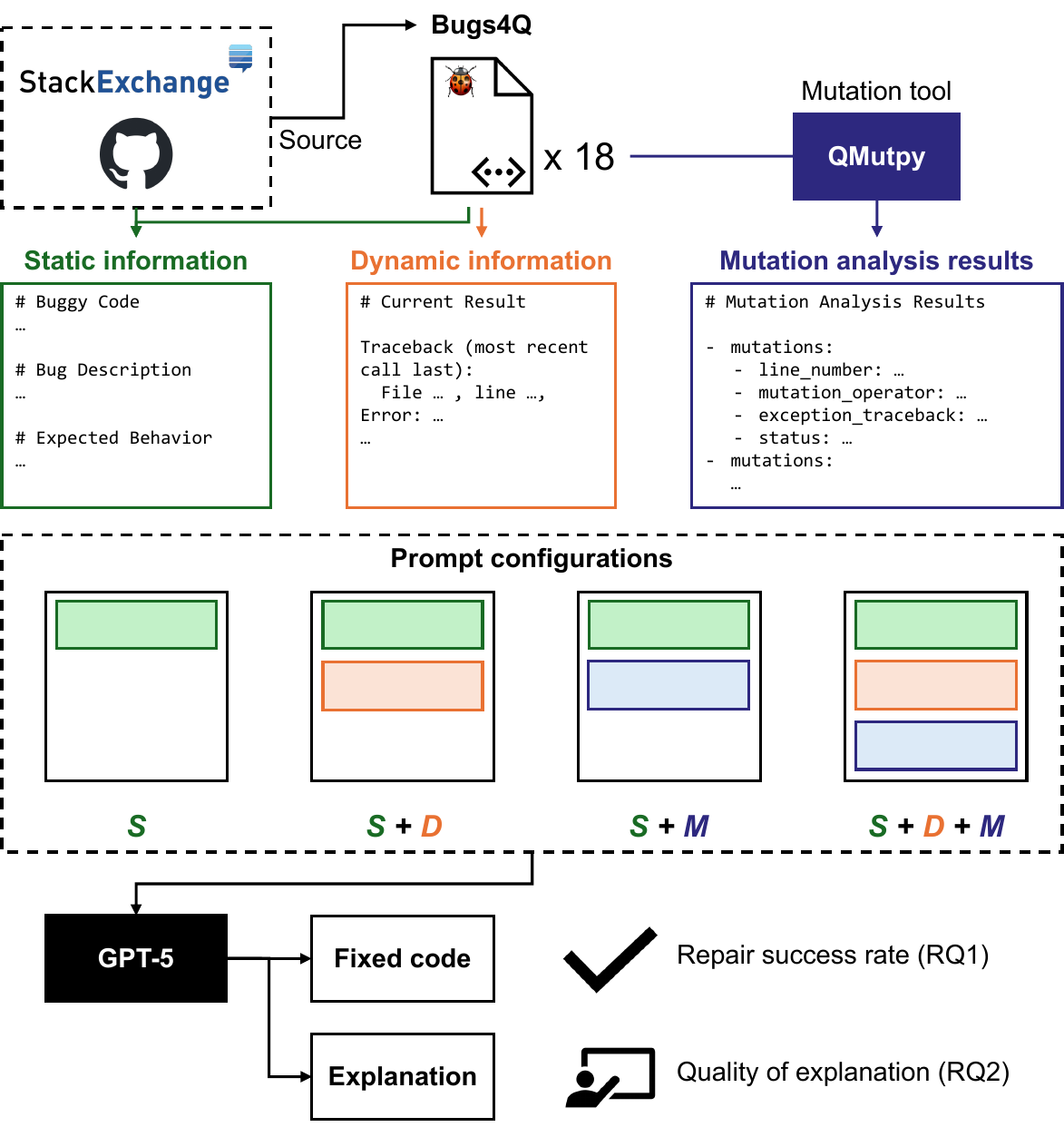}
    \caption{Overview of the experiment.}
    \label{fig:overview}
\end{figure}

This section describes the design of our study, including the benchmark, prompt configurations, and LLMs.
\fig{fig:overview} provides an overview of our experimental process.
Our goal is to investigate how incorporating mutation analysis results into LLM-based repair affects the repair success rate and the quality of generated explanations.  
At a high level, for each buggy program, we prepare four different prompt configurations.
The LLM is instructed to generate both the fixed code and an explanation of the fix regardless of the prompting configuration used.

\subsection{Bug Benchmark}

In this study, we use Bugs4Q~\cite{zhao2023jss}, a benchmark of real-world bugs in quantum programs implemented using Qiskit~\cite{qiskit}.
It contains 42 buggy programs and their fixed versions, collected from GitHub, Stack Overflow, and Stack Exchange.
We used Bugs4Q benchmark because it satisfied the following criteria:
\begin{enumerate}
  \item It contains bugs found in real-world quantum programs written by human developers.
  \item It includes test scripts to verify the behavior of both the buggy and fixed code.
\end{enumerate}
The first criterion is necessary to compare the LLM's explanations of the generated patches with the developers' true intents behind the repairs.
The second criterion is essential to perform mutation analysis and to verify whether the generated patches are correct.


Among the 42 programs in Bugs4Q, we selected 18 programs as our experimental targets.
First, we cloned the replication package of Bugs4Q\footnote{\url{https://github.com/Z-928/Bugs4Q-Framework}} and confirmed whether each bug could be reproduced.
Bugs4Q bugs exhibit two types of symptoms: \textit{Throw Exception  (TE)} (\eg, an exception raised by the Qiskit library) and \textit{Wrong Output (WO)} (\eg, an AssertionError triggered when the measured quantum state differs from the expected state).
19 programs were excluded because their bugs could not be reproduced.
We also excluded 5 programs that did not satisfy either of the following criteria necessary to construct the prompts used in our experiment:
\begin{enumerate}
  \item At least one mutant must be generated when performing mutation analysis.
  \item The URL of the source repository must be accessible.
\end{enumerate}

\subsection{Prompt Construction} \label{subsec: prompt}
To investigate how the information included in prompts influences APR performance for quantum programs, we design the following four prompt configurations (see \fig{fig:overview}):
\begin{itemize}
  \item \textit{S}: Static information only.
  \item \textit{S+D}: Static and dynamic information.
  \item \textit{S+M}: Static information and mutation analysis results.
  \item \textit{S+D+M}: Static information, dynamic information, and mutation analysis results.
\end{itemize}
We describe how each type of information is obtained below.

\noindent \textbf{Static Information.}
We collected three types of static information:
(1) \textit{buggy code} obtained from Bugs4Q;
(2) \textit{bug descriptions} obtained from the source URLs; and
(3) \textit{expected behavior} taken from the source URLs, describing the intended correct program behavior.
(2) and (3) were manually collected by two authors, who inspected the source URLs of each bug and extracted the necessary information. Any disagreements were resolved through discussion to reach a consensus.

\noindent \textbf{Dynamic Information.}
Dynamic information consists of the execution results of buggy programs, including the error message and the accompanying stack trace.

\noindent \textbf{Mutation Analysis Results.}
We applied QMutPy~\cite{fortunato2022tqe} to each buggy program to obtain the mutation analysis results.
QMutPy implements 20 classical and 5 quantum mutation operators, including insertion, deletion, and replacement of quantum gates, as well as insertion and deletion of quantum measurements.
Each mutation operation is applied to a single location at a time.
The mutation operators are applied repeatedly, producing multiple mutated programs (\ie, mutants) for each quantum program.
The original program's test suite is executed on all mutants to determine their outcomes.
We refer to the resulting collection of data as the mutation analysis results, which includes the following information for each applied mutation operation:

\begin{itemize}
  \item \textit{line\_number:} The line number where the mutation operation was applied.
  \item \textit{mutation\_operator:} The name of the mutation operator applied to the program.
  \item \textit{exception\_traceback:} The traceback output when a test execution for the mutant fails.
  \item \textit{status:} The result of the mutation test, categorized into one of the following:
  (1) \textit{killed}, at least one test produces a different outcome;
  (2) \textit{survived}, no change in test outcome;
  (3) \textit{incompetent}, non-viable mutant (e.g., compile error, crash);
  (4) \textit{time\_out}, execution exceeded the time limit (e.g., infinite loop).
\end{itemize}


The system prompt is shared across all four prompt configurations.
It provides detailed instructions on how to interpret and use each piece of information included in the prompt, as well as the strict output format to follow.
All prompts used in our study are available in our replication package~\cite{repurl}.

\subsection{LLM}
In this study, we used the state-of-the-art \textit{GPT-5} via the OpenAI API with default settings.
Because LLMs are stochastic and may produce different outputs for the same input, we generated outputs five times for each prompt configuration.
Consequently, we obtained a total of 18 programs × 4 prompt configurations × 5 generations = 360 generated repairs and their corresponding explanations.

\subsection{Evaluation Metrics}
\noindent \textbf{(RQ1) Repair Success Rate.}
We count a repair as successful if the generated code passes all the tests provided in Bugs4Q.
The repair success rate is defined as the proportion of the 18 programs for which at least one of the five repair attempts was successful.

\noindent \textbf{(RQ2) Quality of Explanation.}
The LLM-generated explanations may describe the cause of the bug, the location of the repair, and the rationale behind the repair.
To evaluate these explanations, we adopt three core elements of patch explanations~\cite{liang2019issre}:
(1) \textit{Position} — where the bug occurs;
(2) \textit{Cause} — why the bug occurs;
(3) \textit{Change} — how the bug is repaired.
The quality of explanations is evaluated using three criteria proposed in previous work~\cite{nauta2023acmcs,sobania2023ssbse}:
\begin{itemize}
  \item \textit{Correctness:} Is the explanation accurate with respect to the ground-truth fixed code?
  \item \textit{Completeness:} Does the explanation fully describe all changes in the patch?
  \item \textit{Complexity:} Does the explanation contain unnecessary complexity?
\end{itemize}

For each of the three explanation elements (Position, Cause, Change), we evaluate whether it satisfies each of the three criteria (Correctness, Completeness, Complexity) through a binary judgment (yes/no). 
For example, given an LLM-generated explanation, we evaluate the \emph{Position} element by answering the following three questions: 1) does the explanation \emph{correctly} explain the position of the bug?, 2) is the explanation \emph{complete} and mentions all locations that were modified for the repair?, and 3) does the explanation include unnecessary \emph{complexity} to explain the location of the bug?

Evaluating the three elements of patch explanation (Position, Cause, Change) therefore resulted in 9 (3 elements × 3 criteria) total binary evaluations per explanation generated by the LLM. We only considered the first generated repair patch among the five produced for each prompt, resulting in a total of 72 explanations.
Two authors independently assessed each explanation, resolving any disagreements through discussion.

\section{Results}
We ran all experiments on simulators on classical computers due to the limited availability of quantum hardware and the innate instability of quantum computing posing a challenge for the reproducibility of our experiment.

\subsection{RQ1: \RQone}

\begin{table}[b]
  \centering
  \caption{Repair success rates by prompt configuration and bug type.
  WO: Wrong Output, TE: Throw Exception.
  The maximum value in each column is highlighted in bold.
}
  \label{tab:rq1-result}
  \renewcommand{\arraystretch}{1}
  \begin{tabular}{lccc}
    \toprule
    Prompt & Total [\%] & WO [\%] & TE [\%] \\
    \midrule
    \textit{S}  & 77.8 & 70.0                        & \textbf{87.5} \\
    \textit{S+D}  & 88.9 & 90.0                      & \textbf{87.5} \\
    \textit{S+M}  & 83.3 & 80.0                      & \textbf{87.5} \\
    \textit{S+D+M}  & \textbf{94.4} & \textbf{100.0} & \textbf{87.5} \\
    \bottomrule
  \end{tabular}
\end{table}
\tab{tab:rq1-result} shows the repair success rates for each prompt configuration and bug type.
The ``Total'' column reports the repair success rate across all 18 programs, while the ``TE'' and ``WO'' columns show those for each bug type (eight and ten programs, respectively). Out of 18 programs, 17 of them could be repaired by at least one prompt configuration.

\textbf{Mutation analysis results are most effective when used in combination with dynamic information.}
Among all prompt configurations, \textit{S+D+M} achieved the highest repair success rate at 94.4\% in total.
When considering only WO bugs, \textit{S+D+M} successfully repaired all programs.
Interestingly, \textit{S+M} achieved a lower total repair success rate than \textit{S+D}.
Because mutation analysis requires running tests, the availability of \textit{M} implies that \textit{D} is also available.
Thus, the lower success rate of \textit{S+M} is not concerning, as \textit{S+D+M} can easily be constructed in such cases.
In contrast, for TE bugs, the set of successfully repaired programs (seven out of eight) was identical across all prompt configurations.
This indicates that dynamic information and mutation analysis results have no impact on buggy code in this category.
For WO bugs, the program executes without crashing, making runtime information such as mutation analysis results helpful for repair.
For TE bugs, the program terminates with an error, making the bug evident and static information alone sufficient.

\summarybox{Answer to RQ1}{
Combining dynamic information with mutation analysis results is most effective, yielding a total repair success rate of 94.4\%.
It also achieved a 100\% repair success rate for the bug type of the Wrong Output.
}

\subsection{RQ2: \RQtwo}

Table~\ref{tab:rq2-result} reports how many of the 18 programs satisfy each criterion for each prompt configuration.
The two evaluators agreed on 79.2\% of the evaluations (513 out of 648 = 72 explanations × 9 evaluation items), and Cohen’s $\kappa$~\cite{cohen1960kappa} was $0.48$, indicating moderate agreement~\cite{landis1977agreement}.
This value is comparable to the $\kappa = 0.55$ reported in prior work~\cite{kang2024fse} evaluating LLM-generated explanations for fault localization.

\renewcommand{\arraystretch}{1.1}{
\begin{table}[h]
  \centering
  \caption{Quality of explanation for each prompt configuration. $\uparrow$ ($\downarrow$) indicates that a higher (lower) value is better.
  The best value in each row is highlighted in bold.
  }
  \label{tab:rq2-result}
  \begin{tabular}{llllll}
    \toprule
    Criterion & Element & \textit{S} & \textit{S+D} & \textit{S+M} & \textit{S+D+M} \\
    \midrule
    \multirow[t]{3}{*}{Correctness ($\uparrow$)} & Position & 12 & 12 & 13 & \textbf{14} \\
                                                 & Cause & \textbf{14} & 13 & 12 & 12 \\
                                                 & Change & 7 & \textbf{9} & 5 & 7 \\
    \midrule
    \multirow[t]{3}{*}{Completeness ($\uparrow$)} & Position & 13 & \textbf{15} & 12 & \textbf{15} \\
                                                  & Cause & \textbf{18} & 15 & 16 & 15 \\
                                                  & Change & 13 & \textbf{16} & 13 & \textbf{16} \\
    \midrule
    \multirow[t]{3}{*}{Complexity ($\downarrow$)} & Position & 2 & 2 & \textbf{1} & \textbf{1} \\
                                                  & Cause & 2 & 2 & \textbf{1} & \textbf{1} \\
                                                  & Change & 8 & 7 & 8 & \textbf{6} \\
    \bottomrule
  \end{tabular}
\end{table}
}

\textbf{From the bolded values in Table~\ref{tab:rq2-result}, \textit{S+D+M} achieved the best scores in six out of nine evaluation items, indicating that this prompt configuration produces better explanations compared to other prompts.}
In particular, for the position element, \textit{S+D+M} achieved the best scores across all criteria, indicating that this prompt configuration is especially effective at positional explanations, leveraging dynamic information and mutation analysis results.
On the other hand, \textit{S+D+M} is less effective for the cause element. 
For this element, the best scores in correctness and completeness were achieved by the prompt configuration \textit{S}. 
One possible reason is that most of the information relevant to the cause of the repair is already contained in the static information.
Mutation analysis results are more effective for detailed positional explanations than for explaining the cause of the bug.

\summarybox{Answer to RQ2}{
Mutation analysis results combined with dynamic information achieved the best scores in six out of nine evaluation items by making positional descriptions more accurate, complete, and concise.}

\section{Discussion}

\subsection{Unique Successful Repairs Across Prompt Configurations}
\begin{figure}[h]
    \centering
    \includegraphics[width=0.85\linewidth]{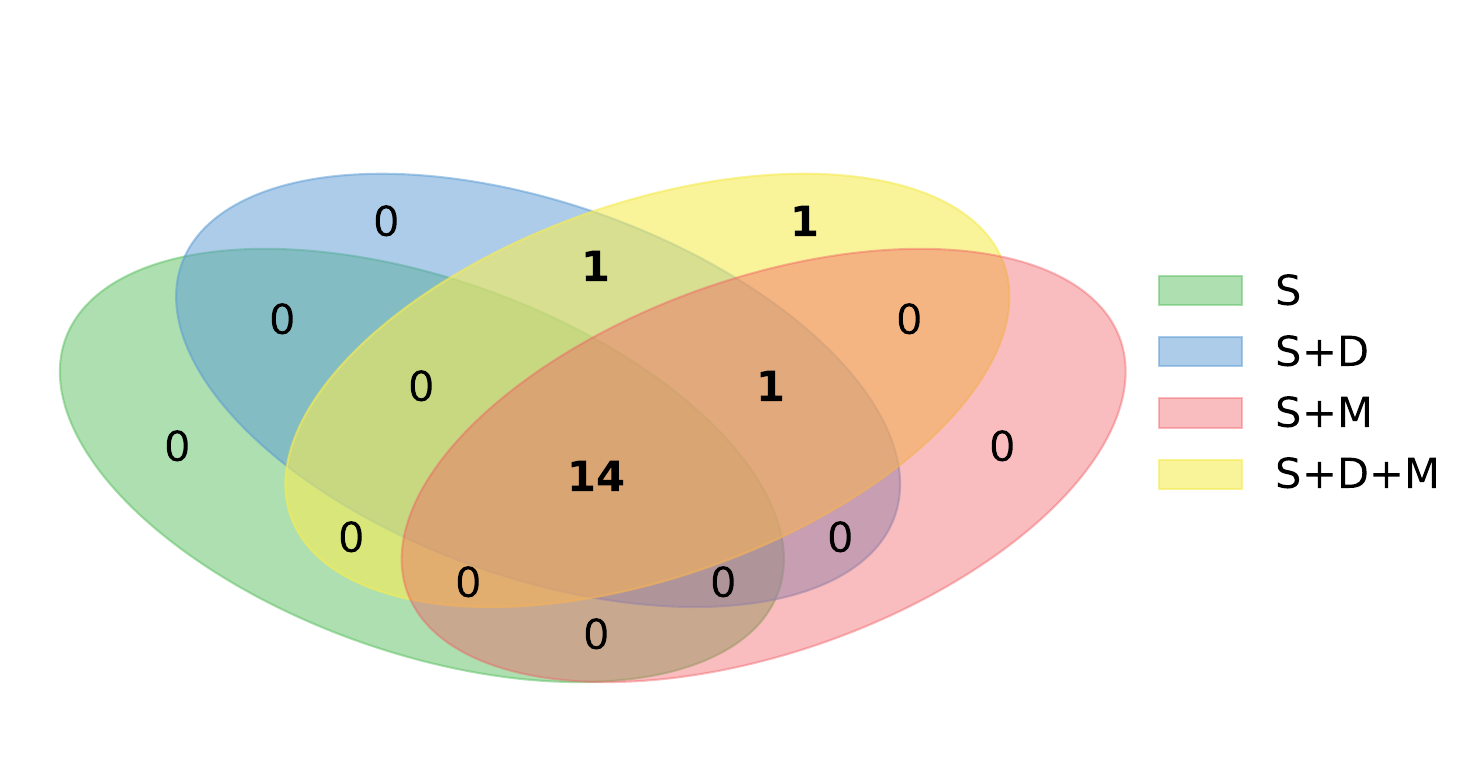}
    \caption{Successful repairs by prompt configuration.}
    \label{fig:venn_diagram}
\end{figure}

Since RQ1 showed the effectiveness of dynamic information and mutation analysis results, we further investigate which programs particularly benefit from these components.
Figure~\ref{fig:venn_diagram} shows the number of programs fixed by each prompt configuration.
While all configurations repair 14 of the 18 buggy programs, the \textit{S+D+M} configuration is the only one that succeeds in repairing 17 programs.
This indicates that dynamic information and mutation analysis results did not hinder repair effectiveness, as the \textit{S+D+M} configuration encompasses all successfully repaired programs. Our results demonstrate that mutation analysis can enhance the effectiveness of LLM-based APR for real-world buggy quantum programs, providing a concrete benefit.

\subsection{Future Research Plan}
\noindent \textbf{(1) Exploring Additional Contextual Information.}
One limitation of our study is that it applies only to quantum programs for which mutation analysis can be performed.
For three programs, we were unable to generate any mutants even though the mutation tool execution was successful.
To address this limitation, we plan to incorporate additional types of contextual information (e.g., relevant sections of the Qiskit documentation) alongside mutation analysis results.

\noindent \textbf{(2) Identifying Domain Knowledge Gaps.}
As another direction for future work, we plan to explore how LLM-generated explanations can help developers identify gaps in their quantum-specific knowledge.
A prior survey shows that contributors to quantum open-source software (OSS) often lack sufficient understanding of physics and quantum computing~\cite{shaydulin2020icsew}.
Our workflow could help developers recognize such gaps through explanations generated during the repair.
To validate the practical utility of this approach, we plan to conduct surveys with OSS developers.
While LLM-based explanations of quantum algorithms have shown promise~\cite{d2024esem}, explanations derived from program repair represent an unexplored avenue.
\section{Threats to Validity}

\noindent \textbf{Construct Validity.}
In RQ1, we defined a successful repair as one that passes all tests, but this does not guarantee that the repair fully matches the developer’s intended modification.
This may affect the validity of the reported repair success rate.

\noindent \textbf{Internal Validity.}
Due to the stochastic nature of GPT-5, the same prompt does not necessarily yield identical results.
Although we generated five outputs for each prompt configuration, the inherent variability of LLM outputs may still influence the findings.
In RQ2, two authors manually evaluated the quality of the explanations to mitigate subjectivity.
However, some degree of subjective judgment is unavoidable, and misclassifications may have occurred due to limited domain knowledge in quantum programming.

\noindent \textbf{External Validity.}
This study was conducted using quantum programs on the Qiskit simulator, and it is unclear whether similar results would be obtained on actual quantum computers. 
Since our investigation is limited to a single benchmark (Bugs4Q) and a single quantum framework (Qiskit), a full validation using a wider range of experimental subjects (\ie, different bug benchmarks, other frameworks, or LLMs other than GPT-5) is needed.
\section{Conclusion}
In this study, we demonstrated the effectiveness of incorporating mutation analysis results into prompts for LLM-based APR of quantum programs using real-world bug benchmarks.
From our experiment, we find that the \textit{S+D+M} prompt configuration, which includes static information, dynamic information, and mutation analysis results, achieves the highest repair success rate of 94.4\%. 
We also find that explanations generated using \textit{S+D+M} exhibit the lowest complexity while providing accurate, complete, and concise descriptions of bug positions. 
These results indicate that combining mutation analysis results with dynamic information can improve both the repair success rate and the quality of LLM-generated explanations.
This work presents the first empirical evidence that mutation analysis can serve as valuable context information for improving LLM-based APR of quantum programs. 
These results point toward new directions for developing APR techniques for quantum programs that enhance both reliability and explainability.


\section{Data Availability}

All data, benchmarks, scripts, and prompts used in this study are publicly available in our replication package~\cite{repurl}.

\section*{Acknowledgment}
We gratefully acknowledge the financial support of: (1) Japan Society for the Promotion of Science, Grant-in-Aid for Scientific Research (B), Grant Number JP25K03102, (2) Japan Society for the Promotion of Science, Grant-in-Aid for Scientific Research (A), Grant Number JP24H00692, (3) Japan Society for the Promotion of Science, Grant-in-Aid for Scientific Research (B), Grant Number JP23K24823, (4) Japan Science and Technology Agency (JST) as part of Adopting Sustainable Partnerships for Innovative Research Ecosystem (ASPIRE), Grant Number JPMJAP2415, and (5) the Inamori Research Institute for Science for supporting Yasutaka Kamei via the InaRIS Fellowship. 

\bibliographystyle{IEEEtran}
\bibliography{reference}

@misc{qiskit,
      title={Quantum computing with {Q}iskit},
      author={Javadi-Abhari, Ali and Treinish, Matthew and Krsulich, Kevin and Wood, Christopher J. and Lishman, Jake and Gacon, Julien and Martiel, Simon and Nation, Paul D. and Bishop, Lev S. and Cross, Andrew W. and Johnson, Blake R. and Gambetta, Jay M.},
      year={2024},
      doi={10.48550/arXiv.2405.08810},
      eprint={2405.08810},
      archivePrefix={arXiv},
      primaryClass={quant-ph}
}

@inproceedings{chen2023icse,
  title={The smelly eight: An empirical study on the prevalence of code smells in quantum computing},
  author={Chen, Qihong and C{\^a}mara, R{\'u}ben and Campos, Jos{\'e} and Souto, Andr{\'e} and Ahmed, Iftekhar},
  booktitle={Proceedings of the IEEE/ACM 45th International Conference on Software Engineering},
  pages={358--370},
  year={2023}
}

@article{openja2022jss,
  title={Technical debts and faults in open-source quantum software systems: An empirical study},
  author={Openja, Moses and Morovati, Mohammad Mehdi and An, Le and Khomh, Foutse and Abidi, Mouna},
  journal={Journal of Systems and Software},
  volume={193},
  pages={111458},
  year={2022}
}

@inproceedings{shaydulin2020icsew,
  author = {Shaydulin, Ruslan and Thomas, Caleb and Rodeghero, Paige},
  title = {Making Quantum Computing Open: Lessons from Open Source Projects},
  year = {2020},
  booktitle = {Proceedings of the IEEE/ACM 42nd International Conference on Software Engineering Workshops},
  pages = {451–455},
  numpages = {5}
}

@book{nielsen2010quantum,
  title={Quantum computation and quantum information},
  author={Nielsen, Michael A and Chuang, Isaac L},
  year={2010},
  publisher={Cambridge university press}
}

@article{zhao2023jss,
  title={Bugs4Q: A benchmark of existing bugs to enable controlled testing and debugging studies for quantum programs},
  author={Zhao, Pengzhan and Miao, Zhongtao and Lan, Shuhan and Zhao, Jianjun},
  journal={Journal of Systems and Software},
  volume={205},
  pages={111805},
  year={2023},
  publisher={Elsevier}
}

@article{fortunato2022tqe,
  title={Mutation testing of quantum programs: A case study with Qiskit},
  author={Fortunato, Daniel and Campos, Jose and Abreu, Rui},
  journal={IEEE Transactions on Quantum Engineering},
  volume={3},
  pages={1--17},
  year={2022},
}

@inproceedings{mendiluze2021ase,
  title={Muskit: A mutation analysis tool for quantum software testing},
  author={Mendiluze, E{\~n}aut and Ali, Shaukat and Arcaini, Paolo and Yue, Tao},
  booktitle={Proceedings of the 36th IEEE/ACM International Conference on Automated Software Engineering},
  pages={1266--1270},
  year={2021},
}

@inproceedings{ali2021icst,
  title={Assessing the effectiveness of input and output coverage criteria for testing quantum programs},
  author={Ali, Shaukat and Arcaini, Paolo and Wang, Xinyi and Yue, Tao},
  booktitle={Proceedings of the 14th IEEE Conference on Software Testing, Verification and Validation},
  pages={13--23},
  year={2021},
}

@inproceedings{ishimoto2024apsec,
  title = {An Empirical Study on Self-Admitted Technical Debt in Quantum Software},
  author = {Ishimoto, Yuta and Nakamura, Yuto and Katsube, Ryota and Sato, Naoto and Ogawa, Hideto and Kondo, Masanari and Kamei, Yasutaka and Ubayashi, Naoyasu},
  booktitle = {Proceedings of the 31st Asia-Pacific Software Engineering Conference},
  pages = {41--50},
  year = {2024}
}

@inproceedings{zappin2025icse,
  title={When Quantum Meets Classical: Characterizing Hybrid Quantum-Classical Issues Discussed in Developer Forums},
  author={Zappin, Jake and Stalnaker, Trevor and Chaparro, Oscar and Poshyvanyk, Denys},
  booktitle={Proceedings of the IEEE/ACM 47th International Conference on Software Engineering},
  pages={2931--2943},
  year={2025},
}

@article{ishimoto2025arxiv,
  title={Evaluating Mutation-based Fault Localization for Quantum Programs},
  author={Ishimoto, Yuta and Kondo, Masanari and Ubayashi, Naoyasu and Kamei, Yasutaka and Katsube, Ryota and Sato, Naoto and Ogawa, Hideto},
  journal={arXiv preprint arXiv:2505.09059},
  year={2025}
}

@article{tan2025fse,
  title={HornBro: Homotopy-Like Method for Automated Quantum Program Repair},
  author={Tan, Siwei and Lu, Liqiang and Xiang, Debin and Chu, Tianyao and Lang, Congliang and Chen, Jintao and Hu, Xing and Yin, Jianwei},
  journal={Proceedings of the ACM on Software Engineering},
  volume={2},
  number={FSE},
  pages={734--756},
  year={2025},
  publisher={ACM New York, NY, USA}
}

@inproceedings{guo2024qse,
  title={On repairing quantum programs using ChatGPT},
  author={Guo, Xiaoyu and Zhao, Jianjun and Zhao, Pengzhan},
  booktitle={Proceedings of the 5th ACM/IEEE International Workshop on Quantum Software Engineering},
  pages={9--16},
  year={2024}
}

@article{li2024tosem,
  title={Automatic repair of quantum programs via unitary operation},
  author={Li, Yuechen and Pei, Hanyu and Huang, Linzhi and Yin, Beibei and Cai, Kai-Yuan},
  journal={ACM Transactions on Software Engineering and Methodology},
  volume={33},
  number={6},
  pages={1--43},
  year={2024},
  publisher={ACM New York, NY}
}

@article{murillo2025tosem,
  title={Quantum software engineering: Roadmap and challenges ahead},
  author={Murillo, Juan Manuel and Garcia-Alonso, Jose and Moguel, Enrique and Barzen, Johanna and Leymann, Frank and Ali, Shaukat and Yue, Tao and Arcaini, Paolo and P{\'e}rez-Castillo, Ricardo and Garc{\'\i}a-Rodr{\'\i}guez de Guzm{\'a}n, Ignacio and others},
  journal={ACM Transactions on Software Engineering and Methodology},
  volume={34},
  number={5},
  pages={1--48},
  year={2025},
  publisher={ACM New York, NY}
}

@article{leite2025tosem,
  title={Testing and debugging quantum programs: The road to 2030},
  author={Leite Ramalho, Neilson Carlos and Amario de Souza, Higor and Lordello Chaim, Marcos},
  journal={ACM Transactions on Software Engineering and Methodology},
  volume={34},
  number={5},
  pages={1--46},
  year={2025},
  publisher={ACM New York, NY}
}

@misc{chatgpt,
  title        = {{ChatGPT}},
  author       = {{OpenAI}},
  year         = {2022},
  howpublished = {\url{https://chat.openai.com/}},
  note         = {Accessed: 2025-11-04}
}

@misc{gpt5,
  title        = {{Introducing gpt-5}},
  author       = {{OpenAI}},
  year         = {2025},
  howpublished = {\url{https://openai.com/index/introducing-gpt-5/}},
  note         = {Accessed: 2025-11-12}
}

@article{mendiluze2025ese,
  title={Quantum circuit mutants: Empirical analysis and recommendations},
  author={Mendiluze Usandizaga, E{\~n}aut and Ali, Shaukat and Yue, Tao and Arcaini, Paolo},
  journal={Empirical Software Engineering},
  volume={30},
  number={3},
  pages={100},
  year={2025},
  publisher={Springer}
}

@article{nauta2023acmcs,
  title={From Anecdotal Evidence to Quantitative Evaluation Methods: A Systematic Review on Evaluating Explainable AI},
  author={Nauta, Meike and Trienes, Jan and Pathak, Shreyasi and Nguyen, Elisa and Peters, Michelle and Schmitt, Yasmin and Schl\"{o}tterer, J\"{o}rg and van Keulen, Maurice and Seifert, Christin},
  journal={ACM Comput. Surv.},
  volume={55},
  number={13s},
  pages={42},
  year={2023},
  publisher={Association for Computing Machinery}
}

@inproceedings{liang2019issre,
  title={How to Explain a Patch: An Empirical Study of Patch Explanations in Open Source Projects},
  author={Liang, Jingjing and Hou, Yaozong and Zhou, Shurui and Chen, Junjie and Xiong, Yingfei and Huang, Gang},
  booktitle={Proceedings of the 2019 IEEE 30th International Symposium on Software Reliability Engineering},
  pages={58-69},
  year={2019}
}

@article{le2019commacm,
  title={Automated program repair},
  author={Le Goues, Claire and Pradel, Michael and Roychoudhury, Abhik},
  journal={Communications of the ACM},
  volume={62},
  number={12},
  pages={56--65},
  year={2019},
  publisher={ACM New York, NY, USA}
}

@article{zhang2023tosem,
  title={A survey of learning-based automated program repair},
  author={Zhang, Quanjun and Fang, Chunrong and Ma, Yuxiang and Sun, Weisong and Chen, Zhenyu},
  journal={ACM Transactions on Software Engineering and Methodology},
  volume={33},
  number={2},
  pages={1--69},
  year={2023},
  publisher={ACM New York, NY}
}

@inproceedings{jin2023fse,
  title={Inferfix: End-to-end program repair with llms},
  author={Jin, Matthew and Shahriar, Syed and Tufano, Michele and Shi, Xin and Lu, Shuai and Sundaresan, Neel and Svyatkovskiy, Alexey},
  booktitle={Proceedings of the 31st ACM joint european software engineering conference and symposium on the foundations of software engineering},
  pages={1646--1656},
  year={2023}
}

@inproceedings{bouzenia2025icse,
  title={RepairAgent: An Autonomous, LLM-Based Agent for Program Repair},
  author={Bouzenia, Islem and Devanbu, Premkumar and Pradel, Michael},
  booktitle={Proceedings of the IEEE/ACM 47th International Conference on Software Engineering},
  pages={694--694},
  year={2025}
}

@artic{ehsani2025arxiv,
      title={Hierarchical Knowledge Injection for Improving LLM-based Program Repair}, 
      author={Ramtin Ehsani and Esteban Parra and Sonia Haiduc and Preetha Chatterjee},
      year={2025},
      journal={arXiv preprint arXiv:2506.24015}
}

@inproceedings{xia2024issta,
  title={Automated program repair via conversation: Fixing 162 out of 337 bugs for \$0.42 each using chatgpt},
  author={Xia, Chunqiu Steven and Zhang, Lingming},
  booktitle={Proceedings of the 33rd ACM SIGSOFT International Symposium on Software Testing and Analysis},
  pages={819--831},
  year={2024}
}

@inproceedings{luo2022saner,
  title={A comprehensive study of bug fixes in quantum programs},
  author={Luo, Junjie and Zhao, Pengzhan and Miao, Zhongtao and Lan, Shuhan and Zhao, Jianjun},
  booktitle={Proceedings of the 2022 IEEE International Conference on Software Analysis, Evolution and Reengineering},
  pages={1239--1246},
  year={2022},
  organization={IEEE}
}

@misc{google_best_practice,
  author       = {Google},
  title        = {Google best practice},
  url = {https://quantumai.google/cirq/google/best_practices},
  note         = {Accessed: 2025-11-13},
  year         = {2025}
}

@inproceedings{sobania2023ssbse,
  title={Evaluating explanations for software patches generated by large language models},
  author={Sobania, Dominik and Geiger, Alina and Callan, James and Brownlee, Alexander and Hanna, Carol and Moussa, Rebecca and L{\'o}pez, Mar Zamorano and Petke, Justyna and Sarro, Federica},
  booktitle={Proceedings of the International Symposium on Search Based Software Engineering},
  pages={147--152},
  year={2023},
  organization={Springer}
}

@article{cohen1960kappa,
  title={A coefficient of agreement for nominal scales},
  author={Cohen, Jacob},
  journal={Educational and psychological measurement},
  volume={20},
  number={1},
  pages={37--46},
  year={1960},
  publisher={Sage Publications Sage CA: Thousand Oaks, CA}
}

@article{landis1977agreement,
  title={The measurement of observer agreement for categorical data},
  author={Landis, J Richard and Koch, Gary G},
  journal={biometrics},
  pages={159--174},
  year={1977},
  publisher={JSTOR}
}

@article{kang2024fse,
  title={A quantitative and qualitative evaluation of LLM-based explainable fault localization},
  author={Kang, Sungmin and An, Gabin and Yoo, Shin},
  journal={Proceedings of the ACM on Software Engineering},
  volume={1},
  number={FSE},
  pages={1424--1446},
  year={2024},
  publisher={ACM New York, NY, USA}
}

@inproceedings{d2024esem,
  title={Exploring LLM-driven explanations for quantum algorithms},
  author={d'Aloisio, Giordano and Fortz, Sophie and Hanna, Carol and Fortunato, Daniel and Bensoussan, Avner and Mendiluze Usandizaga, E{\~n}aut and Sarro, Federica},
  booktitle={Proceedings of the 18th ACM/IEEE International Symposium on Empirical Software Engineering and Measurement},
  pages={475--481},
  year={2024}
}

@misc{repurl,
  title        = {Leveraging Mutation Analysis for LLM-based Automated Quantum Program Repair},
  author       = {Anonymous},
  year         = {2025},
  publisher    = {Zenodo},
  doi          = {10.5281/zenodo.17626083},
  url          = {https://doi.org/10.5281/zenodo.17626083}
}

\end{document}